\documentclass[aip]{revtex4-1}

\usepackage{amssymb,amsmath,amsfonts,amstext,mathtools}

\def\ep{\epsilon}
\def\Ga{\Gamma}
\def\om{\omega}
\def\vph{\varphi}
\def\th{\theta}

\def\bfb{\mathbf{B}}
\def\bfme{\mathbf{E}}
\def\bfa{\mathbf{a}}
\def\bfml{\mathbf{L}}
\def\bfmx{\mathbf{X}}
\def\bfz{\mathbf{z}}
\def\bfe{\mathbf{e}}

\def\bfp{\mathbf{p}}
\def\bfmn{\mathbf{N}}
\def\bfr{\mathbf{r}}
\def\bfq{\mathbf{q}}
\def\bfw{\mathbf{w}}

\def\sfml{\mathsf{L}}
\def\sfb{\mathsf{b}}
\def\sfa{\mathsf{a}}
\def\sfc{\mathsf{c}}

\def\caln{\mathcal{N}}
\def\calr{\mathcal{R}}

\def\p{\partial}
\def\na{\nabla}

\def\bq{\begin{equation}}
\def\eq{\end{equation}}
\def\bqy{\begin{eqnarray}}
\def\eqy{\end{eqnarray}}

\def\ncr{ \nonumber \\ }

\def\upind{_}
\def\x{\times}
\DeclarePairedDelimiter\norm{\lVert}{\rVert}
\def\avg{\texttt{avg}}
\def\osc{\texttt{osc}}

\def\tensa{\bar{\bar{\bfa}}}
\def\bfbz{\bar\bfz}

\def\sfab{\bar{\mathsf{a}}}
\def\sfcb{\bar{\mathsf{c}}}

\def\ubu{\bar{\sfc}\sfb'\sfc}
\def\ubr{\bar{\sfc}\sfb'\sfa}
\def\rbu{\bar{\sfa}\sfb'\sfc}
\def\rbr{\bar{\sfa}\sfb'\sfa}

\def\ubb{\bar{\sfc}\sfb'\sfb}
\def\rbb{\bar{\sfa}\sfb'\sfb}
\def\bpb{\sfb'\sfb}

\begin{document}

\title{A gyro-gauge independent minimal guiding-center reduction
\\
by Lie-transforming the velocity vector field}

\author
{
L.~de~Guillebon}

\email{de-guillebon@cpt.univ-mrs.fr}

\author
{M.~Vittot
}

\affiliation{Centre de Physique Th\'eorique, Aix-Marseille Universit\'e, CNRS, CPT, UMR 7332, 13288 Marseille, France}

\affiliation{Universit\'e de Toulon, CNRS, CPT, UMR 7332, 83957 La Garde, France}

\begin{abstract}
We introduce a gyro-gauge independent formulation of a simplified guiding-center reduction, which removes the fast time-scale from particle dynamics by Lie-transforming the velocity vector field. This is close to Krylov-Bogoliubov method of averaging the equations of motion, although more geometric. At leading order, the Lie-transform consists in the generator of Larmor gyration, which can be explicitly inverted, while working with gauge-independent coordinates and operators, by using the physical gyro-angle as a (constrained) coordinate. This brings both the change of coordinates and the reduced dynamics of the minimal guiding-center reduction order by order in a Larmor radius expansion. The procedure is algorithmic and the reduction is systematically derived up to full second order, in a more straightforward way than when Lie-transforming the phase-space Lagrangian or averaging the equations of motion. The results write up some structures in the guiding-center expansion. Extensions and limitations of the method are considered. 
\end{abstract}

\keywords{guiding-center, Lie-transform, averaging, fast time-scale, Larmor radius expansion, gauge-independent coordinates.}

\maketitle

%----------------------------------------------

	\section{Introduction}
	\label{intro}

 Particle dynamics in a strong static magnetic field implies a separation of scales that allows for a reduction, the so-called guiding-center reduction, which removes the fast time-scale from the dynamics, builds a constant of motion, the magnetic moment, and thus provides a slow motion reduced by two dimensions \cite{CaryBriz09}. The fast time-scale relies on the gyro-angle, which is an angle measuring the Larmor gyration. The main principle is to remove the presence of the gyro-angle in the dynamics of the other coordinates, by building suitable coordinates order by order in a small parameter expansion, related to the Larmor radius. The averaging transformation is usually identified by Lie-transforming the phase-space Lagrangian, because it naturally provides the constant of motion and a Hamiltonian structure for the slow reduced motion. 
 
 Guiding-center theory is widely used in plasma physics \cite{Bitt04, GoldRuth10}, because many physical phenomena rely on time scales much slower than the gyro-frequency, and spatial scales much larger than the Larmor radius. For instance, gyrokinetics is a kinetic model of plasma dynamics induced by guiding-center theory, and it is a key model in the study of magnetically confined hot plasmas, where it has proved efficient for the study of plasma micro-turbulence \cite{BrizHahm07}. 
 
 The usual coordinate for the gyro-angle suffers from several issues, both from a mathematical and from a physical point of view \cite{Little88, Sugi08, Krom09, Sugi09, BurbQin12}: it is gauge-dependent, does not exist globally for a general magnetic geometry, and induces a disagreement between the coordinates and the physical state, since it implies an anholonomic momentum. In order to avoid these issues and to give a more intrinsic framework to the theory, we consider performing the reduction using the initial physical gauge-independent coordinate for the gyro-angle. 
 
 This coordinate is constrained, which makes the scheme more involved. As a first step towards the full reduction, we study here a simplified guiding-center reduction, where we focus on the averaging reduction. The advantage is that only the four slow reduced coordinates have to be changed, since only they have to be gyro-averaged. For the remaining coordinates one can rely on the initial physical coordinates, which are gauge-independent. The scheme also is strongly simplified, since one can work directly on the equations of motion, which makes it easier to identify how to deal with the gyro-angle using an intrinsic but constrained coordinate. \\

 This minimal guiding-center transformation is obtained here by Lie-transforming directly the velocity vector field, which is the geometric point of view of how a coordinate transformation affects the equations of motion \cite{CaryLie, Little82}. As a result, it is much more efficient than previous methods either Lie-transforming the phase-space Lagrangian \cite{CaryBriz09, Little83} or averaging the equations of motion \cite{Krus62, BogoZuba55, BogoMitr61, NortAdia, NortRome78}. This makes it easier to go to higher order in the reduction and to identify structures in guiding-center expansions. The whole second-order reduction is obtained in a direct computation, despite the complicated expressions of the second-order transformation, and the algorithm is so simple that it can be very easily implemented in a computer to reach any order. 
 
 The method is not intended to replace the traditional derivation by Lie-transforming the phase-space Lagrangian. Indeed, with some additional refinements, it can also provide the constant of motion \cite{GuilMagMom}, but it does not guarantee a Hamiltonian structure for the slow reduced coordinates,  with all the associated conservation properties. All the same, it induces a complementary and faster derivation to get additional information on the reduction, and especially it is the first step towards a gauge-independent approach of guiding-center theory. \\

 When applied to the velocity vector field, the Lie-transform method is close to Krylov-Bogoliubov method \cite{KrylBogo47}, but more geometric and more straightforward. Bogoliubov's procedure \cite{BogoZuba55, BogoMitr61} also relied on a near-identity change of coordinates in the equations of motion and achieved greater efficiency than other methods working on the equations of motion, but it relied on substitutions and chain rules, whose process remained rather involved, so that it was not implemented further than the first-order transformation. The Lie-transform method condensates all the procedure into one single operation, related to the transformation generator. 

  It also comprises an interesting approach of the guiding-center arbitrariness. Indeed, the guiding-center reduction is not unique, as is again emphasized by some recent results \cite{ParrCalv11, BrizTron12}. Such an arbitrariness caused troubles in some works dealing with the equations of motion, where it was unclear what the natural choices were to be at second order \cite{NortRome78}. Hamiltonian approaches \cite{Little83, Little79, Little81} did not really address the arbitrariness in the averaging procedure, because they average not just the dynamics of the four reduced coordinates, but the whole reduced Lagrangian, with its seven components, including the Hamiltonian and the Poisson bracket. Now, when Lie-transforming the equation of motion, the arbitrariness is distinct and the natural choice consists in setting to zero all arbitrary terms in the transformation generator, exactly as it was done at first order to define the guiding-center position. This also agrees with Bogoliubov's works \cite{BogoZuba55, BogoMitr61}. 
 
 The resulting guiding-center reduction is unique and minimal. The requirements are just tailored to reduce the dynamical dimensions, which means just to extract a slow dynamics for the four reduced coordinates, the guiding-center position and the parallel velocity or the pitch-angle (the angle between the magnetic field and the particle velocity). Then, the method is forthright from the requirements to the related equations to be solved; it just corresponds to making a near-identity change of coordinates, i.e. a Lie-transform of the velocity vector field. Computations are straightforward and fully algorithmic, since the equations are solved just by an explicit inversion of the Larmor gyration generator. Last, the result is the minimal transformation that allows for the desired reduction; especially, the vector field generating the transformation is a pure gyro-fluctuation. \\

 The paper is organized as follows. In sect. 2, the goals of the reduction are introduced and written as equations for the reduced coordinates. It is shown that these equations can be solved explicitly at any order in the Larmor radius, and the iteration mechanism of the reduction is identified. In sect. 3, it is shown how the reduction is computed and the resulting algorithm is given. In sect. 4, the result is written to full second order in the Larmor radius, obtained in a straightforward computation; a comparison is done with previous guiding-center reductions, and some insights are brought into the general structure of guiding-center results. 
 
 In sect. 5, a few possible extensions and limitations of the method are studied, such as the presence of an electric field or non minimal guiding-center reductions. For instance, averaging also the dynamics of $\theta$ is considered, as well as including the magnetic moment in the reduced coordinates or providing the reduced dynamics with a Hamiltonian structure.

	\section{Lie-transforming the velocity vector field}
 	\label{p1}

 The dynamical system is simply a charged particle with position $\bfq$, momentum $\bfp$, mass $m$ and charge $e$, under the influence of a static inhomogeneous magnetic field $\bfb$. The motion is given by the Lorentz force
\begin{align}
	\dot \bfq &= \tfrac {\bfp}{m}
	\,,\ncr
	\dot \bfp &= \tfrac {\bfp}{m}\x e\bfb
	\,.\notag
\end{align}

 For the sake of clarity, we consider no electric field. The presence of an electric field satisfying the guiding-center ordering would not change the method (see sect. 5). 
 
 When the magnetic field is strong, the motion implies a separation of time-scales. This is best seen by choosing convenient coordinates for the momentum space
\begin{align}
	p &:= \norm \bfp
	\,,\ncr
	\vph &:= \arccos \left(\tfrac{\bfp\cdot\sfb}{\norm{\bfp}}\right)
	\,,
	\label{ChgVar}
	\\
	\sfc &:= \tfrac{\bfp_\perp}{\norm{\bfp_\perp}}
	\,,\notag
\end{align}
where $\sfb:= \tfrac{\bfb}{\norm{\bfb}}$ is the unit vector of the magnetic field, and $\bfp_\perp:=\bfp- (\bfp\cdot\sfb)\sfb$ is the orthogonal projection of the momentum onto the plane perpendicular to the magnetic field. The coordinate $p$ is the norm of the momentum, $\vph$ is the so-called pitch-angle, i.e. the angle between the velocity and the magnetic field. Following Littlejohn's notations \cite{Little83, Little79, Little81}, the vector $\sfc$ is the unit vector of the perpendicular velocity.

 Differentiating formulae (\ref{ChgVar}), the equations of motion are found as 
\begin{align}
	\dot \bfq &= \tfrac {\bfp}{m}
	\,,\ncr
	\dot p & = 0
	\,,\ncr
	\dot \vph & = -\tfrac{\bfp}{m}\cdot\na\sfb\cdot\sfc
	\,,\ncr
	\dot \sfc & = - \tfrac{e B}{m}\sfa 
	- \tfrac{\bfp}{m}\cdot\na\sfb\cdot(\sfc\sfb + \sfa\sfa \cot \vph )
	\,,\notag
\end{align}
where $\bfp$ is now a shorthand for $p(\sfb \cos \vph + \sfc \sin \vph)$, $B$ is the norm of the magnetic field and $\sfa:=\sfb\x\sfc$ is the unit vector of the Larmor radius, following Littlejohn's notations, so that $(\sfa,\sfb,\sfc)$ is a (rotating) right-handed orthonormal frame. The computation of $\dot\sfc$ is rather involved, but it can be avoided by using formula (\ref{GradC}).\\

 In the case of a strong magnetic field, the only fast term, the Larmor frequency $\om_L:=\tfrac{e B}{m}$ concerns only one coordinate, $\sfc$, the direction of the vector $\bfp_\perp$ in the $2$-dimensional plane perpendicular to the magnetic field. This corresponds to an angle, the so-called gyro-angle, and measures the Larmor gyration of the particle momentum around the magnetic field. 

 To get a true scalar angle instead of the vector $\sfc$, one chooses at each point $\bfq$ in space a direction which will be considered as the reference axis $\bfe_1(\bfq) \in \bfb^\perp(\bfq)$. Then, the gyro-angle $\theta$ is defined from the oriented angle between the chosen reference axis $\bfe_1(\bfq)$ and the vector $\sfc$:
\bq
	\sfc = - \sin \theta \bfe_1 - \cos \theta \bfe_2
	\,,
\label{DefThetaPerp}
\eq
and the equation of motion for $\theta$ is
$$
	\dot \theta = 
	\tfrac{e B}{m} 
	+ \cot \vph \tfrac{\bfp}{m}\cdot\na\sfb\cdot\sfa 
	+ \tfrac{\bfp}{m}\cdot\na\bfe_1\cdot \bfe_2
	\,,
$$
with $\bfe_2:=\sfb\x\bfe_1$ is the unit vector such that $(\sfb,\bfe_1,\bfe_2)$ is a fixed right-handed orthonormal frame.\\

 To emphasize the fast term of the dynamics in strong $B$, the equations of motion can be expanded in $B^{-1}$. Writing $\bfz:=(\bfq,p,\vph,\theta)$, we get 
\begin{align}
	\dot \bfz &= 
			\tfrac{eB}{m}
			\left[
						\left(
				\begin{smallmatrix}
				0 \\ 0 \\ 0 \\ 1
				\end{smallmatrix}
			\right)
			+
			\tfrac{p\sin\vph}{eB}
			\left(
				\begin{smallmatrix}
				\sfb\cot\vph+\sfc \\ 
				\raisebox{1.5ex}{~} 0 \raisebox{-1ex}{~}\\ 
				-(\sfb\cot\vph+\sfc)\cdot\na\sfb\cdot\sfc \\ 
				(\sfb\cot\vph+\sfc)\cdot
					(\cot \vph \na\sfb\cdot\sfa
					+ \na\bfe_1\cdot \bfe_2)
				\end{smallmatrix}
			\right)
			\right]
	\ncr
	& =\dot \bfz_{-1} + \dot\bfz_0 
	\,,\ncr
	\text{with~} \dot\bfz_{-1} &:= 
			\left(
				\begin{smallmatrix}
				0 \\ 0 \\ 0 \\ \tfrac{e B}{m}
				\end{smallmatrix}
			\right)
	\,,\ncr
	\dot\bfz_0 &:= 
			\left(
				\begin{smallmatrix}
				\tfrac {\bfp}{m} \\ 
				\raisebox{1.5ex}{~} 0 \raisebox{-1ex}{~}\\ 
				-\tfrac{\bfp}{m}\cdot\na\sfb\cdot\sfc \\ 
				\tfrac{\bfp}{m}\cdot
					(\cot \vph \na\sfb\cdot\sfa
					+ \na\bfe_1\cdot \bfe_2)
				\end{smallmatrix}
			\right)
	\,,
\label{DefOrdMotion}
\end{align}
where the indices correspond to the order in $B^{-1}$. The term of order $B^{1}$ has dimension of $\tfrac{eB}{m}$, whereas all the terms of order $B^0$ have dimension of $\tfrac{\bfp}{m}\cdot\na$, if we use the correspondence between the vector field $\dot\bfq$ and the differential operator $\dot\bfq\cdot\na$. 

 The ordering parameter, that is the ratio of two following terms in the perturbation expansion, is then $\ep=\tfrac{\dot\bfz_{0}}{\dot\bfz_{-1}}=\tfrac{p\sin\vph}{eB} \na$, which is the dimensionless quantity $r_L\na$, with $r_L:=\tfrac{p \sin\vph}{eB}$ the Larmor radius. This parameter is well-known as the magnetic inhomogeneity at the scale of the Larmor radius, because in $r_L\na$, the gradient always acts on the magnetic field, which is the only local property of the configuration space. 
 
 The ordering parameter is sometimes considered just as the Larmor radius $r_L$, or as $B^{-1}$, in agreement with an expansion in strong magnetic field. Also, it is often considered symbolically as $e^{-1}$, because it is an equivalent expansion but it avoids to deal with a space-dependent parameter involving $B$; this is symbolic, because $e^{-1}$ is not dimensionless. Notice that the ordering parameter $r_L\na$ is not just a scalar but an operator; it has only a kind of dimensional meaning: the terms of order $\ep^2$ may not have a prefactor of $(r_L\na)^2$, but they will have the dimension of $\tfrac{p^2}{e^2 B^2}\na^2$. This point will be illustrated by the results of sect. 4. \\

 The goal is to isolate the dynamics of the slow variables from the fast variable, i.e. to perform a change of coordinates $\tau: \bfz\rightarrow \bfbz$ such that the dynamics of the remaining coordinates $(\bar\bfq,\bar\vph,p)$ does not depend on $\theta$. This is already obtained for $p$, so one only has to change coordinates on $\bfq$ and $\vph$. The coordinate transformation $\tau$ transfers to functions by duality through the "push-forward" operator $\mathsf T^{-1}$, defined by the scalar invariance property \cite{CaryBriz09}:
 $$
 	(\mathsf T^{-1}f)(\bar \bfz)= f(\tau^{-1} \bar \bfz)\,.
$$

 At the lowest order in the Larmor radius $\ep^{-1}$, the requirements are trivially satisfied. So, the transformation can be near-identity. It can be written as the exponential of a Lie-transform $\bfbz = e^{-\bfmx}\bfz$, with $-\bfmx$ a vector field, generator of the diffeomorphism $\tau$, which satisfies 
$$
	\bfmx_{p,\th}=0
	\,,
$$
since only the coordinates $\bfq$ and $\vph$ need to be changed. The index notation is used to indicate the components, e.g. $\bfmx_\vph$ denotes the component $\vph$ of the vector $\bfmx$, and $\bfmx_{p,\th}$ denotes a $2$ dimensional vector, whose coordinates are the components $p$ and $\th$ of $\bfmx$. Through the transformation, the equations of motion become
$$
	\dot \bfz \rightarrow \dot \bfbz 
	= e^{\sfml } \dot \bfz
	\,,
$$
with $\sfml:=\bfml_\bfmx$ the Lie-transform along the vector field $\bfmx$. 

Now, the goal is that the equations of motion for $\bfq$ and $\vph$ do not depend on $\theta$, which means that all their non-zero Fourier components (i.e. purely oscillatory terms) are zero: 
$$
	\osc (\dot \bfbz)_{\bfq,\vph} 
	= 0
	\,,
$$
where following Littlejohn's notations, $\osc=1-\avg$ is the projector onto gyro-fluctuations, with $\avg$ the complementary projector onto gyro-averages:
$$
	\avg (f) = \tfrac{1}{2 \pi}\int_0^{2\pi}\! d\theta~ f
$$
for any function $f$.

 Last, $\sfml $ is expanded in series in the small parameter $\ep$ 
\bq
	0 
	= \osc \left( \dot \bfbz \right)_{\bfq,\vph} 
	= \osc \left( e^{\sfml_1+\sfml _2+...} \dot \bfz \right)_{\bfq,\vph}
	\,,
\label{EqToSolve}
\eq
where $\sfml_i$ is of order $\ep^i$. \\

 Equation (\ref{EqToSolve}) is the equation to be solved for the change of coordinates $\sfml_{n+1}$. Expanding it in series in $\ep$, we get an equation for each order:
\begin{align}
	0 & = \osc \left( \dot\bfz_{-1} \right)_{\bfq,\vph}
	\,,\ncr
	0 & = \osc \left( \sfml_1 \dot \bfz_{-1} + \dot\bfz_0 \right)_{\bfq,\vph}
	\,,\ncr
	0 & = \osc \left( \sfml_2 \dot \bfz_{-1} + \tfrac{\sfml_1^2}{2} \dot\bfz_{-1} + \sfml_1 \dot\bfz_0 \right)_{\bfq,\vph}
	\,,\ncr
	& \text{...}
\label{EqByOrders}
\end{align}

 At each order $n\geqslant 1$ in $\ep$, the highest-order unknown $\sfml_{n+1}$ is involved only in one term. Isolating it, the equation writes
\bq
	-\osc(\sfml_{n+1}  {(\dot\bfz_{-1})}_{\bfq,\vph})
	= (\bfmn_{n})_{\bfq,\vph}
	\,,
\label{EquationOrdreN}
\eq
where $\bfmn_{n}$ is a shorthand for all the terms of equation (\ref{EqToSolve}) that are of order $n$ and that do not include the unknown $\sfml_{n+1}$, e.g. $\bfmn_1:=\osc \left( \tfrac{\sfml_1^2}{2} \dot\bfz_{-1} + \sfml_1 \dot\bfz_0 \right)_{\bfq,\vph}$.

 The operator to be inverted is 
\begin{align}
	-\osc & (\sfml_{n+1} (\dot\bfz_{-1}))_{\bfq,\vph}
	\ncr
	& =
	-\osc\left( 
	(\bfmx_{n+1})_i\p_i 
	(\dot\bfz_{-1})_{\bfq,\vph} 
	- (\dot\bfz_{-1})_i \p_i 
	(\bfmx_{n+1})_{\bfq,\vph} 
	 \right)
	 \ncr
	 & =	
	\osc\left( 
	(\dot\bfz_{-1})_\theta \p_\theta (\bfmx_{n+1})_{\bfq,\vph}
	 \right)
	 \ncr
	 & =	
	 \tfrac{e B}{m}
	 \p_\theta (\bfmx_{n+1})_{\bfq,\vph} 
	 \,,
\label{OperToInvert}
\end{align}
where, in the first equality we used the usual formula (\ref{LieDeriv}) for the Lie-transform of a vector field, in the second equality we used $(\dot\bfz_{-1})_{\bfq,\vph}=0$, and in the third equality we used that $\osc~\p_\theta$ is just $\p_\theta$, because $\p_\theta$ takes its values in the gyro-fluctuations, i.e. the non-zero Fourier component in the variable $\theta$. 

 The operator $\om_L\p_\theta$ is the generator of Larmor gyration, with the Larmor frequency as a coefficient. Equation (\ref{EquationOrdreN}) has a solution because the right-hand side $(\bfmn_{n})_{\bfq,\vph}$ is in the range of the operator $\p_\th$, since it is a pure gyro-fluctuation. The operator is easily inverted as 
\bq
	 {(\bfmx_{n+1})}_{\bfq,\vph}
	= \avg\left( {(\bfmx_{n+1})}_{\bfq,\vph} \right) 
	+ \tfrac{1}{\om_L} \int\! d\theta~ (\bfmn_{n})_{\bfq,\vph}
	\,,
\label{Solution}
\eq
where $\avg{(\bfmx_{n+1})}_{\bfq,\vph}$ is a free element in the kernel of $\p_\theta$, that is a free gyro-averaged function. We defined $\int\! d\theta~ (\bfmn_{n})_{\bfq,\vph}$ as the primitive of $(\bfmn_{n})_{\bfq,\vph}$ with zero gyro-average, i.e. $\int\! d\theta~ (\bfmn_{n})_{\bfq,\vph}:=\osc \left(\caln\right)$ with $\caln$ any primitive of $(\bfmn_{n})_{\bfq,\vph}$. \\

 As a result, equation (\ref{EqByOrders}) can be solved to arbitrary order in the small parameter $\ep$ through formula (\ref{Solution}). At each order, one only has to write formula (\ref{EqByOrders}) to the $n$-th order, group into $(\bfmn_{n})_{\bfq,\vph}$ all the terms that depend only on quantities that are already explicitly known, expand the result, and last invert the Larmor gyration operator through formula (\ref{Solution}).  

 This is close to Kruskal's averaging procedure \cite{Krus62}, whose principle was to expand the function defining the motion and average this expansion. Here, we focus on the vector field defining the motion and Lie-transform it so as to make it independent of the fast coordinate. Such a procedure actually does not rest on averaging methods so much; it could rather be related to normal form methods \cite{GuckHolm83}, which aim at giving a simplified form to a vector field, when studying the local dynamics, the stability and possible bifurcations of an equilibrium point. 

 In the guiding-center transformation, at each order, the fluctuating part of $\bfmx_{n+1}$ given by formula (\ref{Solution}) is necessary and sufficient to solve equation (\ref{EqToSolve}). The averaged part is completely free, but also completely useless to solve equation (\ref{EqToSolve}). So, a natural choice is to put it to zero
\bq
	\avg\left( \bfmx_{n+1} \right) = 0
	\,.
\label{NullAverage}
\eq

This makes the transformation unique: the result is the minimal guiding-center reduction, i.e. the only transformation which gives the desired requirements and whose only non-zero components are the fluctuating part of $(\bfmx_{n+1})_{\bfq,\vph}$. \\

 In this guiding-center reduction, the transformation affects only the coordinates $(\bfq,\vph)$. The transformation generator has no gyro-angle component. Now, the corresponding coordinate $\theta$ is not intrinsic. It implies a non-trivial gauge fixing, whose global existence can fail \cite{Sugi08, BurbQin12}. It is not purely physical: in all physical results, what appears is the physical quantity $\sfc$, as is clear in the literature \cite{CaryBriz09, Little83, Little81}. The use of $\theta$ is a detour and should be avoidable. 

 So, we remain with the variable $\sfc$ with its initial definition: the unit vector of the perpendicular velocity
\bq
	\sfc
	:=\tfrac{\bfp_\perp}{\norm{\bfp_\perp}}
	=\tfrac{\bfp-(\bfp\cdot\sfb)\sfb}{p \sin \vph}
	\,.
\label{DefPerp}
\eq

 With this variable, the coordinate space is constrained: the gyro-angle $\sfc$ is not independent of the spatial position, since $\sfc\in\sfb^\perp$, and $\sfb$ depends on $\bfq$. When the $\bfq$ coordinate is changed, the $\sfc$ coordinate cannot be kept unchanged, otherwise it may get out of $\sfb^\perp$.  Differentiating relation (\ref{DefPerp}) with respect to $\bfq$, we find 
\bq
	\na\sfc 
	= - \na\sfb\cdot(\sfc\sfb+\sfa\sfa\cot\vph)
	\,.
\label{NaturConnect}
\eq
This formula can be obtained more easily by noticing that in the change of coordinates $(\bfq,\bfp)\longrightarrow(\bfq,p,\vph,\sfc)$, we have
\begin{align}
	& -\sin\vph\na\vph =\na\cos\vph
	=\na\sfb\cdot\tfrac{\bfp}{p}
	=\na\sfb\cdot\sfc\sin\vph
	\ncr
	& \Rightarrow \na\vph = -\na\sfb\cdot\sfc
	\ncr
	& \Rightarrow 0 = \na\left(\tfrac{\bfp}{p}\right)
	=\na\sfc\sin\vph+\na\sfb\cos\vph
	+\na\vph(-\sfb\sin\vph+\sfc\cos\vph)
	\ncr
	& \rule{10ex}{0pt}= (\na\sfc +\na\sfb\cdot\sfc\sfb)\sin\vph
	+ \na\sfb\cdot(1-\sfc\sfc)\cos\vph
	\ncr
	& \Rightarrow \na\sfc 
	= - \na\sfb\cdot(\sfc\sfb+\sfa\sfa\cot\vph)
	\,. 
	\label{GradC}
\end{align}
in which $\na$ means differentiation with respect to $\bfq$ while keeping $\bfp$ constant, and we used that $1=\sfa\sfa+\sfb\sfb+\sfc\sfc$ and that $\sfb$ is a unit vector, which implies $\na\sfb\cdot\sfb=\na\left(\tfrac{\sfb^2}{2}\right)=0$.

 The action of $\na$ on the vector $\sfc$ must be taken into account through formula (\ref{NaturConnect}) when computing the action of $\dot\bfz_\bfq$ or $(\bfmx_{n+1})_\bfq$, but also when defining the components of $\dot\bfz$: writing $\dot f=\dot \bfz\cdot\p_\bfz f$ in coordinates $\bfz:=(\bfq,p,\vph,\sfc)$ with the property (\ref{NaturConnect}) implies 
\begin{align}
	\dot \bfz &= 
			\tfrac{eB}{m}
			\left[
						\left(
				\begin{smallmatrix}
				0 \\ 0 \\ 0 \\ 1
				\end{smallmatrix}
			\right)
			+
			\tfrac{p\sin\vph}{eB}
			\left(
				\begin{smallmatrix}
				\sfb\cot\vph+\sfc \\ 
				\raisebox{1.5ex}{~} 0 \raisebox{-1ex}{~}\\ 
				-(\sfb\cot\vph+\sfc)\cdot\na\sfb\cdot\sfc \\ 
				0
				\end{smallmatrix}
			\right)
			\right]
	\ncr
	& =\dot \bfz_{-1} + \dot\bfz_0 
	\,,\ncr
	\text{with~} \dot\bfz_{-1} &:= 
			\left(
				\begin{smallmatrix}
				0 \\ 0 \\ 0 \\ \tfrac{e B}{m}
				\end{smallmatrix}
			\right)
	\,,\ncr
	\dot\bfz_0 &:= 
			\left(
				\begin{smallmatrix}
				\tfrac {\bfp}{m} \\ 
				\raisebox{1.5ex}{~} 0 \raisebox{-1ex}{~}\\ 
				-\tfrac{\bfp}{m}\cdot\na\sfb\cdot\sfc \\ 
				0
				\end{smallmatrix}
			\right)
	\,.
	\label{MotionPartIntrins}
\end{align}
 Notice that the $\sfc$ component of $\dot \bfz_0$ is zero, because the term $- \tfrac{\bfp}{m}\cdot\na\sfb\cdot(\sfc\sfb + \sfa\sfa \cot \vph )$ in the dynamics of $\sfc$ comes from $\dot\bfz_\bfq \cdot\na f$ with formula (\ref{NaturConnect}).

	\section{The reduction algorithm}
 	\label{p2}

 In this section, it is shown how the computation proceeds for the minimal guiding-center transformation. To lowest order $\ep^{-1}$, the equation to solve (\ref{EqByOrders}) writes
$$
	0 
	= \osc \left( \dot\bfz_{-1} \right)_{\bfq,\vph}
	\,.
$$

From the definition (\ref{DefOrdMotion}) of $\bfz_{-1}$, this is trivially verified, and it is actually a condition for the near-identity Lie-transform to isolate the fast time-scale, which is possible only because at lowest order, the motion concerns only the fast variable $\sfc$.\\

 To order $0$, the equation to solve (\ref{EqByOrders}) is
$$
	0 
	= \osc \left( \sfml_1 \dot \bfz_{-1} 
	+ \dot\bfz_0 \right)_{\bfq,\vph}
	\,.
$$

The solution for the first-order change of coordinates $\bfmx_1$ is given by equation (\ref{Solution}) with the choice (\ref{NullAverage})
\bq
	{(\bfmx_1)}_{\bfq,\vph}
	= \tfrac{1}{\om_L} \int\! d\theta~ \osc(\dot\bfz_0)_{\bfq,\vph}
	\,.
\label{SolFirstOrd}
\eq

 The spatial component gives the lowest-order Larmor radius, which is often identified with the Larmor radius itself
\bqy
	(\bfmx_1)_\bfq 
	& = & \tfrac{m}{e B} \int\! d\theta~ 
	\osc(\dot\bfz_0)_{\bfq}
	\ncr
	& = & \tfrac{p}{e B} \int\! d\theta~ 
	\osc(\sfb\cos \vph + \sfc \sin\vph)
	\ncr
	& = & \tfrac{p}{e B} \int\! d\theta~ 
	\sfc \sin\vph
	= \tfrac{p \sin\vph}{e B} \sfa
	= r_L \sfa
	\,,
\label{FirstOrdLq}
\eqy 
where the action of $\osc$ and $\int\! d\theta$ can be computed trivially by using formula (\ref{DefThetaPerp}). One can avoid this trick and make the computation with purely intrinsic operators using the results of \cite{GuilMagMom}.

 On another hand, the component $\vph$ of equation (\ref{SolFirstOrd}) gives the expression of the first-order change for the coordinate $\vph$
\bqy
	(\bfmx_1)_\vph 
	& = & \tfrac{m}{e B} \int\! d\theta~ 
	\osc(\dot\bfz_0)_{\vph}
	\ncr
	& = & - \tfrac{p}{e B} \int\! d\theta~ 
	\osc((\sfb\cos \vph + \sfc \sin\vph)\cdot\na\sfb\cdot\sfc)
	\ncr
	& = & - \tfrac{p}{e B} \int\! d\theta~ 
	\left( \cos \vph ~\sfb\cdot\na\sfb\cdot\sfc
	+ 2 \sin\vph \na\sfb:  \tensa _2 \right)
	\ncr
	& = & - \tfrac{p\sin\vph}{e B} 
	\left( \cot \vph ~\sfb\cdot\na\sfb\cdot\sfa
	- \tfrac{1}{2} \na\sfb: \tensa _1  \right)
	\,,
\label{FirstOrdLphi}
\eqy 
with $\tensa_1:=-\tfrac{ \sfa\sfc + \sfc\sfa }{2}$ and $\tensa_2:=\tfrac{ \sfc\sfc-\sfa\sfa }{4}$ the standard dyadic tensors of guiding-center works \cite{CaryBriz09, Little81}. Formulae (\ref{FirstOrdLq})-(\ref{FirstOrdLphi}) agree with the usual fluctuating first-order generator of the guiding-center reduction \cite{CaryBriz09, Little83, Little81}. \\

 To order one, formula (\ref{EqByOrders}) writes
$$
	0 = \osc \left( 
	\sfml_2 \dot \bfz_{-1} 
	+ \tfrac{\sfml_1^2}{2} \dot\bfz_{-1} 
	+ \sfml_1 \dot\bfz_0 
	\right)_{\bfq,\vph}
	\,.
$$

 As usual, it is already solved by formula (\ref{Solution}) with condition (\ref{NullAverage})
\bq
	 {(\bfmx_2)}_{\bfq,\vph}
	= \tfrac{1}{\om_L} \int\! d\theta~ 
	\osc \left( 
	\tfrac{\sfml_1^2}{2} \dot\bfz_{-1} 
	+ \sfml_1 \dot\bfz_0 
	\right)_{\bfq,\vph}
	\,.
\label{SolutionL2}
\eq

 All we have to do is to make the left-hand side of equation (\ref{SolutionL2}) explicit. This is completely algorithmic. The first step is to compute the Lie derivatives using the standard formula  
\begin{align}
	\sfml_n \bfw 
	& = \bfml_{\bfmx_n} (\bfw \upind k \p_k)
\label{LieDeriv}
	\\
	& = 
	\Big( (\bfmx_n)\upind j \p_j \bfw\upind k 
	- \bfw\upind j \p_j (\bfmx_n)\upind k \Big) \p_k
	+ (\bfmx_n)\upind j \bfw\upind m [ \p_j , \p_m ]
	\,,\notag
\end{align} 
for any vector field $\bfw$. Einstein convention is used and repeated indices are implicitly summed. The derivative operator $\p_k$ corresponding to the gyro-angle variable $\sfc$ is $\bfp\cdot\sfb\x\p_\bfp=-\sfa\cdot\p_\sfc$, since it is the generator of Larmor gyration \cite{GuilMagMom}, and is equal to $\p_\th$.

 In formula (\ref{LieDeriv}), the commutator of derivatives $[ \p_j , \p_m ]$ appears because the coordinate $\sfc$ is constrained, it is space-dependent, and the corresponding connection involves the pitch-angle. So, the following commutators are non-zero:
\begin{align}
	[ \na , -\sfa\cdot\p_\sfc ] &= 
	\cot \vph ~(\na\sfb\cdot\sfc) 
	~(\sfa\cdot\p_\sfc)
	\,,\ncr
	[ \na , \p_\vph ]  &= 
	-(1+\cot^2\vph)
	(\na\sfb\cdot\sfa) 
	~(\sfa\cdot\p_\sfc)
	\,, \ncr
	[ \na_i , \na_j ] &= 
	(1+\cot^2\vph) ~\Big[
	(\na_i\sfb\cdot\sfc) 
	~(\na_j\sfb\cdot\sfa)
	\,, \ncr
	& \rule{14ex}{0pt} - (\na_j\sfb\cdot\sfc)
	~(\na_i\sfb\cdot\sfa)  \Big]
	~(\sfa\cdot\p_\sfc)
	\,.\notag
\end{align}

 Formula (\ref{LieDeriv}) is systematically applied to $\sfml_1 \dot\bfz_0 $, then to $\sfml_1 \dot\bfz_{-1} $, and last to $\tfrac{\sfml_1}{2}\left( \sfml_1 \dot\bfz_{-1}\right) $, which appear in the right-hand side of (\ref{SolutionL2}). Although uncomplicated, computations must be very orderly to remain tractable since the products and the Leibniz rule generate many terms from a single expression such as $\tfrac{\sfml_1^2}{2}\dot\bfz_{-1} $ .\\

 The second step is to perform the action of the gyro-integral $\int\! d\theta~\osc$. It acts only on the variables $\sfc$ and $\sfa$, involved in expressions such as 
$$
	\sfa\cdot\na  B ~\sfc\cdot\na\sfb\cdot\sfa 
	= (\sfa\sfc\sfa)\upind {ijk}~ (\na B~\na\sfb)_{ijk}
	\,,
$$  
upon which $\int\! d\theta ~\osc$ operates on the first tensor in the left-hand side by mixing the $\sfc$ and $\sfa$. 

 A way to perform the action of $\int\! d\theta ~\osc$ is to compute its action on the basic tensors $\sfc$, $\sfc\sfc$, $\sfc\sfc\sfc$, etc. and to deduce its action on the other tensors by using cross products with the magnetic field $(\sfb\x)\upind{ij}=\ep_{ikj}\sfb\upind k$; this last operator is usually denoted by $\sfb\upind{ij}$. For instance, $(\sfa\sfc\sfa)$ can be written 
\bq
	(\sfa\sfc\sfa)_{ijk}=\sfb\upind{im}~\sfb_{kn}~(\sfc\sfc\sfc)\upind {mjn}
	\,.
	\label{uruVERSuuu}
\eq

 Last, the action of $\int\! d\theta ~\osc$ on the elementary tensors $\sfc^{\otimes N}$ can be computed by using the intrinsic calculus introduced in \cite{GuilMagMom}, or by introducing a local fixed basis $(\bfe_1, \bfe_2)$ through the change of coordinate: 
\bqy
	\sfc &=& - \sin \theta \bfe_1 - \cos \theta \bfe_2
	\,,\ncr
	\sfa &:=& \cos\th \bfe_1  -  \sin\th \bfe_2
	\,,\notag
\eqy
as is standard in guiding-center reductions. With the fixed basis, the action of $\int\! d\theta ~\osc$ is trivial to compute: $\osc$ just cancels the zeroth Fourier component (gyro-average), and $\int\! d\theta$ is an easy integral. If $N$ is large, the computation by hand may be tedious but it remains trivial with a computer. Then one can come back to the initial basis $(\sfc, \sfa)$ to get $\int\! d\theta\osc(\sfc^{\otimes N})$.

 For instance, the lowest-orders formulae are
\begin{align}
	\int\! d\theta\osc~(\sfc)
	&=
	\sfa
	\,, \ncr
	\int\! d\theta\osc~(\sfc\sfc)
	&=
	\frac{\sfc\sfa+\sfa\sfc}{4}
	\,, 
	\label{GyroIntegral}
	\\
	\int\! d\theta\osc~(\sfc\sfc\sfc)
	&=
	\tfrac{1}{3} \big[\sfa\sfc\sfc+\sfc\sfa\sfc+\sfc\sfc\sfa+2\sfa\sfa\sfa\big]
	\,, \ncr
	\int\! d\theta\osc~(\sfc\sfc\sfc\sfc)
	&=
	\tfrac{1}{32} \Big[5(\sfa\sfc\sfc\sfc+\sfc\sfa\sfc\sfc
	+\sfc\sfc\sfa\sfc+ \sfc\sfc\sfc\sfa) 
	\ncr &\rule{5ex}{0pt} 
	+ 3\big(\sfc\sfa\sfa\sfa+\sfa\sfc\sfa\sfa
	+\sfa\sfa\sfc\sfa+\sfa\sfa\sfa\sfc\big)\Big]
	\,.\notag
\end{align} 
 These are all we need to get the second-order reduction. Higher harmonics could be as easily dealt with. In practical computations, the relations given above can often be simplified by the symmetries of the tensor which $\sfc^{\otimes N}$ is contracted with.
 
 Notice that $\int\! d\theta\osc$ is a linear operator which preserves the order in the fast variable $\sfc$ or $\sfa$. It is a pseudo-inverse for $\partial_\theta$. If we restrict the operator $\partial_\theta$ to the set of tensors that are harmonics of order $n$ in the fast variable, it becomes just a matrix with finite dimension $2^n$. If $n=2N+1$ is odd, the matrix is invertible, and is easily obtained (on a computer for instance) and gives the action of $\int\! d\theta\osc$. As regards even harmonics $n=2N$, the kernel of $\partial_\theta$ is not zero, but it is a complementary space to the range of $\partial_\theta$, as is obvious in Fourier series, for instance. So, after identifying the kernel and the range of this operator, one gets an invertible matrix by restricting the operator to its range. Then, the action of the operator $\int\! d\theta\osc$ is just this invertible matrix on $\text{Range}(\partial_\theta)$ and zero on $\ker(\partial_\theta)$. It is actually a very efficient way to perform the action of the operator $\int\! d\theta\osc$. \\

 After using formulae (\ref{uruVERSuuu}) and (\ref{GyroIntegral}) on each term of $\tfrac{\sfml_1^2}{2} \dot\bfz_{-1} + \sfml_1 \dot\bfz_0$, recombining the various terms and simplifying the result with formula 
\bq
	\sfb\cdot\na B = - B \na\cdot\sfb
	\,,
	\label{FormulaDivB}
\eq
coming from $\na\cdot \bfb=0$, one obtains the second-order generator of the minimal guiding-center transformation $\bfmx_2$, which is given by formulae (\ref{SolutionL2X})-(\ref{SolutionL2phi}) in the next section.

 The result is somehow intricate, especially for $(\bfmx_2)_\vph$, but this was expected for a formula at second order, which even was not reached in usual derivations of guiding-center reductions. Actually, it emphasizes how efficient Lie-transforming the velocity vector field is, since it provides by a straightforward computation such a complicated result.  

 The explicit computation at order $2$ illustrates the method to perform the derivation to arbitrary order. To each order in the Larmor radius expansion, the algorithm consists in writing down formula (\ref{Solution}) and making all the terms explicit. Only two special operations are involved: the Lie-derivatives in the computation of $(\bfmn_{n})_{\bfq,\vph}$, given by formula (\ref{LieDeriv}), and the gyro-integral $\int d\theta~\osc$ computed by (\ref{GyroIntegral}); these are automatic operations. 
 
 If computations soon become tedious, it is only because the numbers of terms rapidly increases, as a result of products in Lie derivatives, of the Leibniz rule for the action of gradients, and of the gyro-integral operator. This is what generates the complicated formula (\ref{SolutionL2phi}) for $(\bfmx_2)_\vph$. But this is no trouble, since computations are completely algorithmic and very direct: formula (\ref{SolutionL2phi}) for $(\bfmx_2)_\vph$ was obtained with no trouble by hand, and computations at higher order can be very easily done by computer-assisted symbolic calculus. 

 The fast growth of the number of terms with the order of expansion raises the question of the convergence of the series, but in perturbation expansions, convergence is not the first issue, even if it is important to be addressed \cite{Little82, BogoMitr61}. In addition, a factorial growth is quite standard and convergence is generally obtained only with methods of accelerated convergence \cite{BogoMitr76}. When convergence is not guaranteed or when non-convergence is proven, the asymptotic expansion, with its truncated perturbation expansions, all the same allows for a strong reduction of the effect of the fast time-scale in the dynamics of $(\bar\bfq,\bar\vph)$, as emphasized in \cite{Krus62}. 

 From this point of view, a simplified derivation, such as the minimal guiding-center reduction, is an interesting way to control the iteration process and to obtain more information on the asymptotic behaviour.

\section{The result to second order}

By applying the procedure introduced in last section, the second-order generator of the minimal guiding-center transformation is obtained
\begin{align}
	(\bfmx_2 & )_\bfq:=
	\left(\tfrac{p\sin\vph}{eB}\right)^2
	\left[
	\sfb \left(
		-2\phi\ubb + \frac{\rbr-\ubu}{8}
		\right)
	\right.
	\label{SolutionL2X}
	\\
	& \rule{10ex}{0pt}
	+ \phi^2\frac{\sfc\sfcb-\sfa\sfab}{4}\bpb
	\left. 
	+ \tfrac{\phi}{8}
	\big(
	4\sfc\rbr+7\sfa\ubr-9\sfa\rbu
	\big)
	\right]
	\,,\ncr
%----------------------------
	(\bfmx_2 & )_\vph:=
	\left(\tfrac{p\sin\vph}{eB}\right)^2
	\left[
	\tfrac{B'}{24 B}
	\big(
	4\sfc\ubu-4\sfc\rbr+5\sfa\ubr +5\sfa\rbu
	\big)
	\right.
	\ncr
	& 
	+ \phi^3 
	\Big\lbrace
	\frac{\sfab\sfab-\sfcb\sfcb}{4}\bpb\bpb
	\Big\rbrace
	\label{SolutionL2phi}
	\\
	& 
	+ \phi^2 \left\lbrace
	-\sfcb\sfb''\sfb\sfb
	+
	\big(
	-\sfcb\sfb'\frac{20\sfc\sfcb+9\sfa\sfab}{8}
	+\sfab\sfb'\frac{7\sfc\sfab-16\sfa\sfcb}{8}
	\big)
	\sfb'\sfb
	\right\rbrace
	\ncr
	& + \phi^1
	\left\lbrace
	\frac{\sfab\sfb''\sfa-\sfcb\sfb''\sfc}{8}\sfb
	+(\rbr-\ubu) \frac{9\rbr+7\ubu}{32}
	\right.
	\ncr
	& \left.
	~~~~~~~~
	-\frac{\sfab\sfab-\sfcb\sfcb}{8}\bpb\bpb
	+(\ubr+\rbu) \frac{3\rbu-5\ubr}{32}
	\right\rbrace
	\ncr
	& + \tfrac{\phi^0}{24} 
	\Big\lbrace
	8\sfcb\sfb''\sfc\sfc-5\sfcb\sfb''\sfa\sfa+11\sfab\sfb''\sfc\sfa
	\ncr
	&
	~~~~+\big(
	-16\ubu\sfcb+10\ubr\sfab-11\rbu\sfab-11\rbr\sfcb
	\big)\sfb'\sfb
	\Big\rbrace
	\Big]
	\,,
	\notag
\end{align}
where $\phi$ is a shorthand for $\cot \vph$. To make expressions easier to read, we used the primed notation for spatial gradients, and the over-bar over a vector $\sfc$ or $\sfa$ means matrix transpose, e.g. $\bar\sfa \sfb''\sfa\sfb'\sfc$ means $\sfc\cdot(\na\sfb)\cdot[(\sfa\cdot\na)(\na\sfb)]\cdot\sfa$. This notation is close to Littlejohn's notations $(\sfa\sfb\sfc)$ in \cite{Little83}, but it is more suited for higher-order derivatives.

 In equation (\ref{SolutionL2X}), the first line is the $\sfb$-component, parallel to the magnetic field, and the following two lines are perpendicular to $\sfb$. In equation (\ref{SolutionL2phi}), the first line contains the terms depending on $\na B$ and the following lines are organized as a polynomial of the pitch-angle, or rather its cotangent $\phi$. Note that the terms depending on $\sfb\cdot\na B$ do not appear in the first line, because they are rewritten using formula (\ref{FormulaDivB}). The expressions involved in such a second-order result can be written in many different but equivalent ways, as already noticed by Northrop and Rome, and a rule for standardizing them is needed for the derivation to be efficient.
 
 If there are many terms, this is especially because each term appears several times, with permutations of $\sfa$ and $\sfc$; these permutations are often condensed into one single tensor. For instance, the last line can be written just 
$$
	\na\sfb:\Pi\cdot(\sfb'\sfb), 
$$
by defining the triadic tensor 
$$
	\Pi:=
	-16\sfc\sfc\sfc +10\sfa\sfc\sfa-11\sfc\sfa\sfa-11\sfa\sfa\sfc
$$
Then the number of terms is strongly reduced, each order in $\phi$ has one or two terms, which shows that the result is not that complicated actually. However, we chose to avoid introducing such intermediate quantities, since they make formulae shorter but less explicit.\\

 The second-order vector field gives the minimal guiding-center change of coordinates to second order
$$
	 \bar \bfz = 
	 \left[ 
	 1 - (\sfml_1) - (\sfml_2) 
	 + \left( \tfrac{\sfml_1^2}{2} \right)
	 + O(\ep^3)
	 \right]
	 \bfz
	 \,,\ncr
$$
where the vector fields $\bfmx_1$ and $\bfmx_2$ generating the Lie-transforms $\sfml_1$ and $\sfml_2$ are given by formulae (\ref{FirstOrdLq}), (\ref{FirstOrdLphi}), (\ref{SolutionL2X}) and (\ref{SolutionL2phi}). 

So, to second order, the Larmor radius is 
\begin{align}
\bfr_L = &(\bfz - \bar \bfz )_\bfq
	 = (\bfmx_1)_\bfq 
	 + (\bfmx_2)_\bfq 
	 - \left( \frac{(\bfmx_1)_{\bfz}\cdot\p_{\bfz} 
	 (\bfmx_1)_\bfq}{2} \right) 
	 + O(\ep^3)
	 \ncr 
	 = &
	 \tfrac{p\sin\vph}{eB}
	\sfa
	\label{BarQOrd2}
	\\
	& + 
	\left(\tfrac{p\sin\vph}{eB}\right)^2
	\left[
	\sfb \left(
		-2\phi\ubb + \frac{5\rbr-\ubu}{8}
		\right)
	\right.
	\ncr
	& \rule{14ex}{0pt}
	+ \sfa \tfrac{B'\sfa}{2B}
	+ \phi^2\tfrac{\bpb}{4}
	+ \phi
	\sfa \big(
		\ubr-\rbu
	\big)
	\bigg]
	 \,.
	\notag
\end{align}

 And to second order, the reduced pitch-angle is 
\begin{align}
	 \bar \vph = &
	\left[ 
	 1 - (\sfml_1) - (\sfml_2) 
	 + \left( \tfrac{\sfml_1^2}{2} \right)
	 + O(\ep^3)
	 \right]
	 \vph
	 \ncr 
	  = &
	 \vph 
	 - \tfrac{p\sin\vph}{eB}
	 \left[  
	 -\phi \rbb - \frac{\ubr+\rbu}{4}
	 \right]
	\label{BarPhiOrd2}
	\\
	 & - \left(\tfrac{p\sin\vph}{eB}\right)^2
	\left[
	\tfrac{B'}{12 B}
	\Big\lbrace
	-6\phi\sfa\rbb
	+ 2\sfc\ubu 
	\right.
	\ncr
	&\rule{17ex}{0pt}
	-2\sfc\rbr+\sfa\ubr+\sfa\rbu
	\Big\rbrace
	\ncr
	& 
	+ \phi^3
	\Big\lbrace
	\frac{\sfab\sfab-\sfcb\sfcb}{4}\bpb\bpb
	\Big\rbrace
	\ncr
	& 
	+ \phi^2 \left\lbrace
	-\sfcb\sfb''\sfb\sfb
	+\left(
	-5\sfcb\sfb'\frac{2\sfc\sfcb+\sfa\sfab}{4}
	+ 3 \sfab\sfb'\frac{\sfc\sfab-2\sfa\sfcb}{4}
	\right)\bpb
	\right\rbrace
	\ncr
	& + \phi^1
	\left\lbrace
	\frac{5\sfab(\sfb'\sfb)'\sfa-\sfcb(\sfb'\sfb)'\sfc}{8}
	+\frac{3\sfab\sfab+\sfcb\sfcb}{8}\bpb\bpb
	\right.
	\ncr
	& \left.
	~~~~~~~~- \frac{3(\rbr-\ubu)^2}{32}
	+(\ubr+\rbu) \frac{\rbu-3\ubr}{16}
	\right\rbrace
	\ncr
	& + \tfrac{\phi^0}{12} 
	\Big\lbrace
	4\sfcb\sfb''\sfc\sfc-\sfcb\sfb''\sfa\sfa+7\sfab\sfb''\sfc\sfa
	\ncr
	& 
	~~~~+
	\big(
	-8\ubu\sfcb+5\ubr\sfab-4\rbu\sfab-7\rbr\sfcb
	\big)\sfb'\sfb
	\Big\rbrace
	\Big]
	\,.\notag
\end{align}

 The ordering in the Larmor radius is obvious. The first line corresponds to the zeroth- and first-order terms, and all the following lines are the second-order terms, which are organized as in formula (\ref{SolutionL2phi}). \\

 We wrote all these formulae exactly as they are yielded by the procedure, because it illustrates both the mechanism of the derivation and the structure of the resulting change of coordinates. 
 
 Indeed, all the above formulae rely only on a very restricted alphabet of entities: $B$, $\sfb$, $\sfc$, $\sfa$, $\na$ and the variable $\vph$, or more precisely $\phi$, if we discard the $\sin \vph$ occurring in the pre-factor (Larmor radius) and the quantities $p$ and $e$, which are mute parameters in the derivation. 
 
 Each formula is a series in the Larmor radius $r_L=\tfrac{p \sin\vph}{eB}$, or more precisely in the ordering parameter of guiding-center reduction $\ep=r_L \na$, i.e. the magnetic inhomogeneity at the scale of the Larmor radius. This dimensionless parameter is not just a number, it is to be understood in the sense that the only dimensional quantities involved are the Larmor radius, appearing as a pre-factor to the power given by the expansion order, and the gradients (acting on the magnetic field $B$ and $\sfb$), with the same order as the expansion order. 
 
 At each order in $\ep$, formulae are polynomials in each variables (except $B$ whose disposition obeys trivial dimensional rules). Using this fact makes the derivation much simpler. Finally, the iteration at each order consists in one single formula (\ref{Solution}) with two elementary operators (derivatives and gyro-integral); in addition, they are applied on terms composed of very few elementary entities, and those terms are just polynomials. 

 As the order in the small parameter $\ep$ grows up, the polynomial increases its order in each of the variables. From the iteration mechanism, a rough estimate shows that the $n$-th-order generator $\bfmx_n$ should be a polynomial of order at most $\phi^{2n-1}\sfb^{3n-1}\sfc^{2n}$, in addition to being a monomial in $r_L^n\na^n$. Such features in the reduction transformation are useful to consider when the asymptotic behaviour is addressed. For instance, a hypothetical convergence condition would clearly involve the expected condition on the magnetic inhomogeneity at the scale of the Larmor radius, but the role of $\phi$ suggests conditions on the pitch-angle as well. This means that in the guiding-center reduction, the direction of the particle velocity must not be too close to the direction of the magnetic field. This is in complete agreement with the physical intuition, but the polynomial behaviour can help make this intuitive statement more precise. 

 Another interesting feature is well emphasized by the formulae as they are written: there is a link of parity between the order in $\phi$, the order in $r_L$ and the fast-angle harmonic (which corresponds to the order in $\sfc$ or $\sfa$). All non-zero terms of order $r_L^i\phi^j$ are harmonics of parity $(-1)^{i+j}$ in $\sfc$ for $\bfmx_\bfq$, and harmonics of parity $(-1)^{i+j+1}$ in $\sfc$ or $\sfa$ for $\bfmx_\vph$. This parity relation could already be observed in the first-order results (\ref{FirstOrdLq})-(\ref{FirstOrdLphi}), although it was not so obvious, because few terms were present.

 It seems that the polynomiality in $\phi$ and the parity relation were first noticed in \cite{GuilMagMom}, where they were observed in the derivation of the magnetic moment series, and originated from the structure of the operator to be inverted for the secular equation. Here, they are obtained in the guiding-center reduction, and appear as originating directly from the equations of motion (\ref{DefOrdMotion}) or (\ref{MotionPartIntrins}), and from the action of derivatives, especially $\p_\vph \phi$ and $\na\sfc$. 
 
 This polynomiality is related to the coordinate $\phi$. Previous works used the parallel velocity $v_\parallel$ as a coordinate instead of $\phi$. Their results were not polynomials (see e.g. \cite{CaryBriz09, Little83, BogoZuba55}), but when expressed with the coordinate $\phi$, they become also polynomials.\\

 The second-order results can be expressed using well-known quantities in guiding-center works. For instance, in the term of order $\phi^1$ in formula equation (\ref{BarPhiOrd2}), $\ubr+\rbu$ and $\ubu-\rbr$ can be recognized, commonly written $-2\na\sfb:\tensa_1$ and $4\na\sfb:\tensa_2$ where the dyadic tensors $\tensa_1=-\tfrac{\sfa\sfc+\sfc\sfa}{2}$ and $\tensa_2=\tfrac{\sfc\sfc-\sfa\sfa}{4}$ are well-known in guiding-center works \cite{CaryBriz09, Little81}, and they were already met in first-order results (\ref{FirstOrdLphi}).

 Also, in the last line of equation (\ref{BarPhiOrd2}), each of the terms is harmonic of order $3$ in the fast angle $\sfc$ or $\sfa$. The sum can be rewritten
\bq
	\big(
	-3 \tau_m \sfab +\sfab\sfb'(\sfb\x)
	-8\na\cdot\sfb ~\sfcb+2\ubr~\sfab
	\big)\bpb
	\,,
\label{MultipleWritings}
\eq
where three out of four terms have been combined to obtain an harmonic of order $1$ in the gyro-angle, which is the maximum we can do. Then, in the coefficients, the divergence of magnetic field lines $\na\cdot\sfb=\ubu+\rbr$ is recovered, as well as their twist
$$
	\tau_m:=\sfb\cdot\na\x\sfb=(\sfb\x\na)\cdot\sfb
	=\rbu-\ubr
	\,,
$$
which quantities are most used in guiding-center theory. 

 Such an expression as (\ref{MultipleWritings}) is useful to get a physical intuition of the terms involved, but it does not emphasize as much the mechanism of the derivation nor the polynomial structure of the results. Furthermore, it is not unique, e.g. it could equally be written 
$$
	\big(
	- 5 \tau_m \sfab -\sfab\sfb'(\sfb\x)
	-6\na\cdot\sfb ~\sfcb-2\ubu~\sfcb
	\big)\bpb
	\,,
$$
or even 
$$
	\big(
	 -13 \tau_m \sfab -9\sfab\sfb'(\sfb\x)
	-8 \sfcb\sfb'+2\rbr\sfcb
	\big)\bpb
	\,,
$$
or (at least) nine other equivalent ways of writing this expression. In all of them, three out of four terms have been combined to obtain harmonics of order $1$ in the gyro-angle, and only the last term remains an harmonic of order $3$. 

 So, we preferred to give the expressions as they come out from the procedure, decomposed in elementary terms in a unique, standardized way, as is needed for an algorithmic procedure, especially when it can generate very many terms.

 Anyway, it is quite common in perturbation theory and well known in guiding-center theory that when the order increases, formulae become messy and complicated, and the physical interpretation of each term loses some of its relevance \cite{NortRome78}. The main point is to standardize the derivation and the involved expressions, to make things tractable and as clear as possible. \\

 Compared to the literature, our fluctuating first-order results are exactly identical to the usual results of guiding-center derivations, e.g. in \cite{CaryBriz09, BogoZuba55, NortRome78, Little81}. This is because this part of the reduction is unique for all guiding-center reductions, since the arbitrariness is only in the average transformation generator. Our second-order results are almost identical to previous results. For instance, the Larmor radius (\ref{BarQOrd2}) is exactly identical to the standard result of \cite{Little81}, except for the term $\phi^2$, which was absent from Littlejohn's result. As for the pitch-angle, the first-order term in (\ref{BarPhiOrd2}) is exactly the same as in \cite{Little81}, except the averaged term, which is absent from our result. 
 
 A difference with previous results is not a surprise because here the minimal guiding-center reduction is considered, and the averaged part of the transformation generator $\bfmx$ have been set to zero. It is just an effect of the guiding-center reduction non-uniqueness, which was well emphasized in \cite{NortRome78} and \cite{BrizTron12}: even between the classical derivations by Lie-transforming the phase-space Lagrangian, such as \cite{Little83, ParrCalv11, BrizTron12, Little81}, various changes can be identified in the choices adopted for the second-order reduction. 

 In previous publications, the second-order pitch-angle generator $(\bfmx_2)_\vph$ was not computed. The works aiming at averaging the equations of motion were not able to reach the second-order transformation \cite{BogoZuba55, NortRome78}. The works using a Lie-transform of the Lagrangian, although more efficient, required quite a lot of algebra to get $(\bfmx_2)_\vph$ and did not reach the full second-order transformation \cite{CaryBriz09, Little81}. Even the recent work \cite{ParrCalv11} explicitly chose not to compute it, even though it aimed at improving the second-order terms and already made quite a lot of computations. Similarly, the work \cite{BrizTron12} uses a special property between the second- and first-order terms to identify the second-order reduced dynamics without computing the second-order transformation $\bfmx_2$. 
 
 These difficulties can be partially explained because when Lie-transforming the Lagrangian, there is some mixing between the orders in $\ep$ of the various quantities \cite{CaryBriz09, Little83}. Thus, some components of the first-order transformation generator $\bfmx_1$ are identified at the first-order analysis of the reduced Lagrangian $\bar\Ga_1$; other components of $\bfmx_1$ are determined at second-order analysis $\bar\Ga_2$, at the same time as some of the components of $\bfmx_2$; and the last components of $\bfmx_1$ are determined at third order $\bar\Ga_3$, at the same time as some components of $\bfmx_2$ and of $\bfmx_3$. It clearly makes the scheme more involved. 
 
 This is very different from what happens when Lie-transforming the equations of motion, where there is no mixing between the order in $\ep$ of the various quantities, and the non-trivial expression for $(\bfmx_2)_\vph$ was obtained by a direct derivation to order two with very limited algebra. \\

 Let us turn now to the guiding-center equations of motion. Equations (\ref{EqToSolve}) and (\ref{EqByOrders}) show that to second order, the drift equations are given by 
\begin{align}
	\dot{\bar\bfz}
	& =
	e^{\sfml } \dot \bfz
	=
	\calr \left( e^{\sfml } \dot \bfz \right)
	\ncr
	& =
	\calr \left[ 
	1 + (\sfml_1) + (\sfml_2) 
	+ \left( \tfrac{\sfml_1^2}{2} \right)
	+ O(\ep^3)
	\right]
	\dot {\bar\bfz}
	\,.
\label{FormMotionReduced}
\end{align}

 This is just the zero Fourier component of $e^{\sfml } \dot \bfz$, which was put to zero when computing the action of $\osc$ in (\ref{EqToSolve}). So, this term was already computed in the derivation. 
 
 In this sense as well, Lie-transforming the equations of motion is more straightforward because one actually directly derives the reduced motion, whereas when Lie-transforming the Lagrangian, one derives the reduced Lagrangian; the reduced equations of motion must be obtained in a second step, for instance by formula (\ref{FormMotionReduced}), which involves some algebra because of the lengthy expressions at order $2$. 
 
 So, the reduced equations of motion are obtained here with no additional computation as 
\begin{align}
	(\dot{\bar\bfz})_\bfq:= &
	\tfrac{p\sin\vph}{m}
	\left[
	\sfb\phi
	+
	\tfrac{p\sin\vph}{eB}  
	\left( 
	\tfrac{B'}{B}\tfrac{\sfc\sfa-\sfa\sfc}{2}
	\right.
	\right.
	\label{DynamReduiQ}
	\\
	& \rule{10ex}{0pt}\left.\left. 
	+\sfb \tfrac{\rbu-\ubr}{2}
	+ \phi^2\big(\sfa\sfcb-\sfc\sfab\big)\sfb'\sfb
	\right)
	\right]
	\ncr
	 = &
	\tfrac{p\sin\vph}{m}
	\Big[
	\sfb\phi
	+
	r_L
	\Big( 
	\tfrac{\sfb\x \na B}{2B} + \sfb\tfrac{\sfb\cdot\na\x\sfb}{2}
	+\phi^2\sfb\x\sfb'\sfb
	\Big)
	\Big]
	\,,\ncr
%--------------------------------
	(\dot{\bar\bfz})_\vph:= &
	\tfrac{p\sin\vph}{m}
	\left[
	-\tfrac{\ubu+\ubr}{2}
	\right.
	\label{DynamReduiPhi}
	\\
	& \rule{5ex}{0pt} \left. +
	\tfrac{p\sin\vph}{eB} \phi 
	\left( 
	\tfrac{B'}{B}\tfrac{\sfa\sfcb-\sfc\sfab}{2}\sfb'\sfb
	+\tfrac{\sfab(\sfb'\sfb)'\sfc-\sfcb(\sfb'\sfb)'\sfa}{2}
	\right)
	\right]
	\ncr
	 = &
	\tfrac{p\sin\vph}{m}
	\left[
	-\tfrac{\na\cdot\sfb}{2}
	+
	r_L\phi
	\left( 
	\tfrac{B'}{2B}\sfb\x\sfb'\sfb + \tfrac{\sfb\cdot\na\x(\sfb'\sfb)}{2}
	\right)
	\right]
	\,.\notag
\end{align}

 In these equations, we first gave the expression provided by the algorithm and then rewrote it using physical quantities, in order to show that the equations are indeed gyro-averages, and do not depend on the fast variables $\sfc$ and $\sfa$.
 
 The right-hand side of equations (\ref{DynamReduiQ})-(\ref{DynamReduiPhi}) is expressed in the reduced variables (e.g. the field is evaluated at the guiding-center position $\bar\bfq$ and the pitch-angle is actually the reduced pitch-angle $\bar \vph$), but we dropped the bars for simplicity. Notice that the right-hand side of equations (\ref{DynamReduiQ})-(\ref{DynamReduiPhi}) is expressed in the initial variables. It is why equations (\ref{SolutionL2X})-(\ref{BarPhiOrd2}) indeed define a coordinate transformation, whereas equations (\ref{DynamReduiQ})-(\ref{DynamReduiPhi}) indeed define the guiding-center dynamics \cite{Little81}. 
 
 In results (\ref{DynamReduiQ})-(\ref{DynamReduiPhi}), the lowest-order term is again exactly identical to the usual guiding-center reductions, and the second-order drifts are almost identical. The guiding-center dynamics (\ref{DynamReduiQ}) contains exactly the four expected terms \cite{BogoZuba55, NortRome78}: the parallel motion along the magnetic field lines, the grad-B drift, the curvature drift, and the Ba\~nos drift. But the Ba\~nos term $\sfb\tfrac{\sfb\cdot\na\x\sfb}{2}$ was not in the reduced dynamics of \cite{Little81}. This is again because the minimal reduction does not aim at minimizing the average dynamics, but at minimizing the change of variables. To make it short, it can be said that the reduction of \cite{Little81} puts the Ba\~nos term into the change of variables $\bfmx$ whereas the minimal reduction lets it in the equations of motion. This difference is mentioned in \cite{BogoZuba55, NortRome78} and further explored in the next section.

	\section{On non-minimal guiding-center reductions}
 	\label{p3}

 In case there is an electric field $\bfme$ in addition to the strong magnetic field, the initial equations of motion (\ref{MotionPartIntrins}) become 
\begin{align}
	\dot \bfz = 
			\tfrac{eB}{m}
			\left[
			\left(
				\begin{smallmatrix}
				0 \\ 0 \\ 0 \\ 1
				\end{smallmatrix}
			\right)
			\right.
		& +
			\tfrac{p\sin\vph}{eB}
			\left.
			\left(
				\begin{smallmatrix}
				\sfb\cot\vph+\sfc \\ 
				\raisebox{1.5ex}{~} 0 \raisebox{-1ex}{~}\\ 
				-(\sfb\cot\vph+\sfc)\cdot\na\sfb\cdot\sfc \\ 
				0
				\end{smallmatrix}
			\right)
			\right.
		\ncr
		&+
			\tfrac{m}{B p\sin\vph}
			\left.
			\left(
				\begin{smallmatrix}
				0 \\ 
				\bfme\cdot\bfp \sin\vph \\ 
				\bfme\cdot\left(\cos \vph ~\tfrac{\bfp}{p} - \sfb\right)  \\ 
				(\bfme\cdot\sfa)\sfa
				\end{smallmatrix}
			\right)
			\right]
	\,,
	\notag
\end{align}
As for the variable $\theta$, its dynamics becomes 
$$
	\dot \theta = \tfrac{e B}{m} 
	+ \cot \vph \tfrac{\bfp}{m}\cdot\na\sfb\cdot\sfa 
					+ \tfrac{\bfp}{m}\cdot\na\bfe_1\cdot\bfe_2
					- \tfrac{e \bfme\cdot \sfa}{p ~ \sin\vph}
	\,.
$$

 In these time evolutions, the two common small parameters of guiding-center theory appear: in strong magnetic field, only one term implies fast dynamics, it is of order $\om_L$; among the other terms, the ones that do not depend on the electric field are of order $\tfrac{p \na}{m}$, and the ones that do depend on $\bfme$ are of order $\tfrac{e ~ \bfme}{p ~ \sin \vph}$. Thus, two ordering parameters are now involved, the magnetic parameter $\ep:=\tfrac{p \na}{\om_L m}=\tfrac{p\na}{e B}$ is the same as in the case of a pure magnetic field; the second parameter $\ep_E:=\tfrac{e ~ \bfme}{\om_L p ~ \sin \vph}=\tfrac{E}{v_\perp  B }$ is induced by the electric field. It is the electric force over the magnetic force, and is assumed to be small in guiding-center theories \cite{CaryBriz09, BrizHahm07, ParrCalv11}. 

 Under this assumption, the procedure for the minimal guiding-center reduction is unchanged; the presence of an electric field only causes some additional contributions in the term $\bfmn_{n}$ of equation (\ref{Solution}), and makes the transformation include one more coordinate: as the norm of the particle momentum is no more conserved, this coordinate must be changed exactly in the same way as the coordinates $(\bfq,\vph)$ in the previous sections. Notice that this minimal guiding-center reduction will imply $\bar p$ to have slow dynamics, but not to be a constant of motion. We turn to this point in a few lines. \\

 In the previous sections, the minimal guiding-center reduction appeared as a natural choice when Lie-transforming the velocity vector field. Other possibilities are available. 
 
 For instance, one can think of treating the gyro-angle $\theta$ as the other coordinates, and changing coordinate $\theta\rightarrow\bar\theta$ to gyro-average its dynamics as well. Thus, all of the reduced dynamics would involve only slow variables. Especially, the reduced gyro-angle $\bar\theta$ would have fast dynamics, but would depend only on the slow dynamics, since it would be independent of $\bar\theta$ itself: $\dot{\bar\theta} = \dot{\bar\theta} (\bar\bfq,\bar\vph,p)$. So, once the motion of the slow variables $( \bar\bfq(t),\bar\vph(t))$ is known, the fast dynamics would be trivial to integrate:
$$
	\bar\theta(t)=\bar\theta(t_0) 
	+ \int_{t_0}^t dt ~\dot{\bar\theta} \Big( \bar\bfq(t),\bar\vph(t),p \Big)
	\,.
$$

 The corresponding requirement relies on equation (\ref{EqToSolve}) written for the gyro-angle component
\bq
	0 
	= \osc \left( e^{\sfml_1+\sfml _2+...} \dot \bfz \right)_{\theta}
	\,,
\label{EqMoyTheta}
\eq
which is to be solved for $\sfml_\theta$ (or rather $\bfmx_\theta$). The operator to be inverted is exactly the same as in the previous sections, since 
\begin{align}
	-\osc & (\sfml_{n+1} (\dot\bfz_{-1}))_{\theta}
	\ncr
	& =
	-\osc\Bigg(
	{(\bfmx_{n+1})}_i \p_i (\dot\bfz_{-1})_\theta
	- (\dot\bfz_{-1})_\theta \p_\theta {(\bfmx_{n+1})}_\theta  
	 \bigg)
	 \ncr
	 & =	
	 \tfrac{e B}{m}
	 \p_\theta {(\bfmx_{n+1})}_{\theta} 
	 + o.t.
	 \,,
	 \label{OperatAvgDotTheta}
\end{align}
where $o.t.$ means other terms that are already known: they do not contain the unknown ${(\bfmx_{n+1})}_{\theta}$ because $(\dot\bfz_{-1})_{\theta}$ does not depend on $\theta$. So, just as in the previous sections, a solution of equation (\ref{EqMoyTheta}) can be found at arbitrary order in the small parameter, and the dynamics of $\theta$ can be averaged as well as the dynamics of $(\bfq,\vph)$, as is generally done in guiding-center reductions \cite{CaryBriz09, BogoZuba55, Little81}.\\

 Notice that the averaging process for the reduced coordinates $(\bfq,\vph)$ is possible without having to deal with resonances \cite{GuckHolm83} because there is no small divisor since the fast part of the dynamics involves only one angle $\theta$, as illustrated by formula (\ref{Solution}). This is a rather broad property. Now, for the dynamics of $\theta$, the averaging process is possible only because the lowest-order term in the motion does not depend on $\theta$ itself, as is illustrated by formula (\ref{OperatAvgDotTheta}). This is a more specific property. 
 
 For instance, after the guiding-center reduction has averaged the motion over the gyro-angle, the bounce-angle can be used to bounce-average the motion \cite{CaryBriz09}. For this reduction, the procedure of section $2$ can  be applied to remove the bounce time-scale from the dynamics of the three other components, but averaging the dynamics of the bounce-angle coordinate $\th_b$ as well is guaranteed only if the lowest-order bounce-angle dynamics is independent of the bounce-angle, which is a condition over the definition of $\th_b$. \\

 A second natural requirement concerns the average part of the change of variables $\avg (\bfmx)$. In the minimal derivation, it is trivially put to zero. In the case of a non minimal reduction, it appears as an additional freedom in the reduction process, which allows for additional requirements, to be chosen. 

 For instance, the transformation being a pure gyro-fluctuation $\avg(\bar\bfz-\bfz)=0$ would be interesting: once the effects of the fast coordinate averaged, then the transformation would become zero; so, the reduced coordinates would remain in some sense close to the initial ones. 
 
 In the minimal guiding-center reduction, this result is almost obtained, since the transformation generator $\bfmx$ is already a pure gyro-fluctuation, but the whole transformation is not a pure fluctuation, because the non-linear terms in $\bfmx$ have non-zero gyro-averages (see, e.g. formulae (\ref{BarQOrd2})-(\ref{BarPhiOrd2})). 
 
 This can be corrected by defining a non-zero average $\avg (\bfmx)$ to cancel the average of the non-linear terms. Indeed, the requirement 
\bq
	0
	=
	\avg\Big(\bar\bfz-\bfz\Big)
	=
	\avg\left(e^{-\sfml}\bfz-\bfz\right)
	\notag
\eq
implies 
\bq
	\avg \Big(\sfml\bfz\Big)
	=
	\avg
	\left( 
		e^{-\sfml}\bfz - (1-\sfml)\bfz
	\right)
	\,.
	\notag
\eq
Now, in the equation at order $n$ 
\begin{align}
	\avg \Big(\bfmx_n\Big)
	&=
	\avg \Big(\sfml\bfz\Big)_n
	=
	\avg
	\Big( 
		e^{-\sfml}\bfz - (1-\sfml)\bfz
	\Big)_n
	\ncr
	&=
	\avg
	\left( 
		\sum_{i=2}^\infty\tfrac{(-\sfml)^i}{i!}
		\bfz
	\right)_n
	\label{ReqTransfFluctu}
	\\
	&=
	\avg
	\left( 
		\tfrac{\sfml_{n-1}\sfml_{1}+\sfml_{n-2}\sfml_{2}+...}{2}\bfz
		-\tfrac{\sfml_{n-2}\sfml_{1}\sfml_{1}+...}{3!}\bfz
		+...
	\right)
	\,,
	\notag
\end{align}
the right-hand side depends only on generators of orders lower than $n$. So formula (\ref{ReqTransfFluctu}) is no more an equation, but just a definition of $\avg (\bfmx)$.  
Up to second order, equation (\ref{ReqTransfFluctu}) writes
\bqy
	\avg (\bfmx_1)&:=& 0
	\,,
	\ncr
	\avg (\bfmx_2)&:=&
	\avg\left( \tfrac{\sfml_1^2}{2} \bfz\right)
	\,.
	\notag
\eqy
The right-hand side was already computed in the minimal derivation, so that the additional requirement does not imply an additional computation. \\

 Another possible requirement, commonly considered, is to include the magnetic moment $\mu$ among the reduced coordinates as a conserved quantity instead of $p$. Indeed, the basic conserved quantity $p$ is not an adiabatic invariant in general when an electric field is present, whereas $\mu$ is so. The requirement is to put to zero not only the fluctuating part of the reduced motion $\osc(\dot\bfbz)_\mu$, as in equation (\ref{EqToSolve}), but its averaged part as well $\avg(\dot\bfbz)_\mu=0$, which simplifies the reduced dynamics in a drastic way. 
 
 For the other coordinates, the same requirement can not be asked, because there is no constant of motion independent of $p$ and $\mu$, since otherwise the Hamiltonian motion with $3$ degrees of freedom and $3$ independent constants of motion would be integrable. For other coordinates, one can not ask so strong a simplification of the reduced dynamics $\avg(\dot\bfbz)$ for all orders in $\ep$, but one can ask it for the orders higher than two for instance. In this way, the reduced motion would be exactly known to all orders even before computing the change of variables. Of course, this is not guaranteed, since it is possible only if the chosen reduced dynamics is equivalent to the initial particle dynamics \cite{GuckHolm83}.\\

 As an example, the minimal guiding-center reduction lets an average contribution (\ref{DynamReduiPhi}) in the equations of motion for the reduced pitch-angle. Perhaps a stronger reduction could cancel it at least for the orders higher than two, by choosing a convenient (non-zero) averaged part $\underline\bfmx$ for the transformation generator. More precisely, in the equation at lowest order, the unknown $\underline\bfmx_1$ has no influence on $\dot{\bar\bfz }_0$, which means that $\bar\vph$ can not be made a constant of motion, as was expected. But at the following order, $\underline\bfmx_1$ now contributes, since at that order, the equation 
$0 = \avg \left( e^{\sfml_1+\sfml _2+...} \dot \bfz \right)_{\vph}$ writes 
\begin{align}
	& \avg \left( 
	\frac{ \bfml_{\underline\bfmx_1}\bfml_{\underline\bfmx_1} 
	+ \bfml_{\widetilde\bfmx_1} \bfml_{\underline\bfmx_1}
	+ \bfml_{\underline\bfmx_1} \bfml_{\widetilde\bfmx_1}  }{2} \dot\bfz_{-1} 
	+ \bfml_{\underline\bfmx_1} \dot\bfz_0 
	\right)_{\vph}
	& =
	o.t.
	\,,
	\label{EquaDiffAverL}
\end{align}
where we used that $\avg \left( \bfml_{\bfmx_2} \dot \bfz_{-1} \right)_\vph =0$; $\underline\bfmx$ and $\widetilde\bfmx$ denote respectively the averaged and the fluctuating part of $\bfmx$; and the symbol $o.t.$ is a shorthand for all the other terms, which do not contain $\underline\bfmx_1$. 

 Equation (\ref{EquaDiffAverL}) shows that, with the additional requirements, the operator to be inverted for $\underline\bfmx$ can be quadratic. Even when it is not, e.g. at the next order, the linear operator may not be trivial; a part of it will be given by $\bfml_{\underline\bfmx_1} \dot\bfz_0 = (\underline\bfmx_1)\upind j \p_j \dot\bfz_0 - (\dot\bfz_0)\upind j \p_j \underline\bfmx_1$, whose coefficients have the non-trivial expression given by (\ref{DefOrdMotion}). Such an operator may not be invertible: if the right-hand side is not in its range. This is the classical problem of secular terms in perturbation theory. Even when the operator is invertible, an explicit inverse may not be obvious to get. 

 Less strong requirements can be considered than putting to zero the higher-order reduced dynamics. For instance, the paper \cite{NortRome78} mentions the possible requirement $\sfb\cdot\dot{\bar\bfq}=0$, i.e. the guiding-center motion is a pure drift across the magnetic field lines, whereas the paper \cite{BogoZuba55} mentions a requirement relating the parallel motion of the guiding-center and the reduced parallel velocity $\bar{v_\parallel}$ (where the parallel velocity $v_\parallel$ is used as a coordinate corresponding to the pitch-angle). In any case, care must be taken about additional requirements. They imply differential equations that are not so simple to deal with, and that even can be impossible to solve. It reminds as well that the minimal guiding-center reduction is very nice with its trivial operator $\tfrac{eB}{m} \p_ \th$ or equivalently $-\tfrac{eB}{m}\sfa\cdot\p_\sfc$. 
 
 The requirement aiming at the magnetic moment plays a special role. Its existence can be viewed as a consequence of the averaging reduction, as shown by Kruskal \cite{Krus62}. The corresponding secular differential equation can be solved at any order, but it is outside the scope of the present paper and will be the topic of another paper \cite{GuilMagMom}. \\

 In the approach based on Lie-transforming the equations of motion, the additional requirements do not change the equation (\ref{EqToSolve}) that gives the fluctuating components of the transformation; they generally aim at a further simplification of the reduced dynamics and imply additional differential equations such as (\ref{EquaDiffAverL}), which impose the averaged components of the transformation. Those equations are not easily solved, and from this point of view, the method of Lie-transforming the phase-space Lagrangian is more powerful, because it does not rely on differential equations, but on algebraic equations, easier to study. 
 
 For instance, the idea to put to zero the averaged components of the reduced motion (at least at orders higher than $2$ or $3$ in the ordering parameter) is included in usual guiding-center derivations, which obtain that it is possible for six components of the Lagrangian, out of seven. This kind of result would not be so easy to get by Lie-transforming the velocity vector field. \\

 Another additional requirement regards the Hamiltonian structure of guiding-center dynamics. The initial motion $\dot\bfz$ is Hamiltonian, and the reduced motion should be Hamiltonian. Indeed, the full reduced motion $\dot\bfbz$ is Hamiltonian, even if it has a different Poisson bracket as the initial motion, since it is just given by a (non-canonical) change of coordinates.
 
 However the true reduced motion involves only the slow variables $(\bar\bfq,\bar\vph)$, and the corresponding dynamical system $(\dot{\bar\bfq},\dot{\bar\vph})$ is not guaranteed to be Hamiltonian, because it is given by a truncation of the full reduced dynamics, and truncations do not preserve the Hamiltonian structure in general. 

 The preservation of the Hamiltonian structure for the $4$-components guiding-center dynamics can be considered as an additional requirement. It is hard to obtain by Lie-transforming the equations of motion, because when deriving reduced models by working on the equations of motion, the Hamiltonian character is not worked on, it is observed a posteriori as preserved or not. On the contrary, a Hamiltonian $4$-components guiding-center dynamics is easily obtained when working on the phase-space Lagrangian, for instance by Lie-transforming it in such a way that the reduced Poisson bracket is quarter-canonical. 
 
 Achieving this in a gauge-independent framework makes the scheme much more involved than the one considered in this paper. The reduction mechanism is much more elaborated, all of the coordinates have to be changed, and the formalism involves not only derivative operators, but also differential forms, which must be considered carefully because of the constrained coordinate $\sfc$. This will be considered in the future.

	\section{Conclusion}

 A gauge-independent minimal guiding-center reduction can be performed at any order in the Larmor radius expansion by Lie-transforming the velocity vector field. The procedure is very efficient and systematic: it just writes the minimal requirements for the reduction, expands the equation in the Larmor radius, and inverts the generator of Larmor gyration to get both the change of coordinates and the reduced equations of motion order by order. The full second-order reduction was straightforwardly obtained, in contrast with previous derivations of the guiding-center reduction. 

 The corresponding transformation generator is a pure gyro-fluctuation and only four of its components are non-zero. This is the bare minimum since exactly the fluctuating part of the slow reduced motion involves a fast time-scale that has to be removed. Thus, the reduction is minimal and unique. All the arbitrary components of the transformation generator are set to zero. \\

 The results bring insights into the structure of the guiding-center formulae, which show up a polynomiality in $\sfb$, $\sfc$, $\sfa$, $r_L$, $\na$ and $\phi$, and a parity relation between the orders in $r_L$ , $\phi$ and the fast angle $\sfc$ or $\sfa$. This polynomiality makes easier the algorithm, which consists in applying at each order two operations (derivatives, and gyro-integration) onto a polynomial of a very restricted alphabet of entities. 
 
 The method can be applied to perform the bounce-average reduction. Also, when an additional electric field is present, the procedure is exactly unchanged even if formulae have additional terms. 
 
 For the gyro-angle, no gyro-gauge was introduced, since the initial physical coordinate $\sfc$ was used. This removes from the theory the issues associated with the traditional gyro-angle $\theta$, and could bring interesting contributions to clarify these issues in previous results. Indeed, a recent work \cite{BurbQin12} was lead to a similar orientation when studying the questions raised by the non-existence of a global gauge to define a gyro-angle. 

 Thus, this approach of the guiding-center reduction, with its minimal procedure and results and its gauge-independence, can contribute to a better understanding of the guiding-center reduction by giving a simplified and more intrinsic point of view. \\

 All the same, it is only the first step towards a gauge-independent guiding-center theory. Standard guiding-center reductions impose additional (non-minimal) requirements for the reduction.  
 
 The magnetic moment is commonly included in the reduced coordinates. This can be done using the gauge-independent gyro-angle and working on the equations of motion by solving the corresponding secular differential equation, but it introduces new features and will be reported elsewhere \cite{GuilMagMom}. 
 
 Some averaged terms can be transferred from the reduced dynamics to the change of variables, in order to get a stronger guiding-center reduction. This means having non-zero gyro-averaged components in the transformation generator, and these components are again obtained by solving secular differential equations. 
 
 Such non-minimal guiding-center reductions are more efficiently obtained by Lie-transforming the phase-space Lagrangian, especially because it mainly relies on algebraic equations. Also, it guarantees a Hamiltonian structure for the $4$-components guiding-center dynamics. Introducing the gauge-independent coordinate in this framework is the next step of the work. It makes the scheme more subtle for the constrained coordinate $\sfc$, especially because of the presence of differential forms.

	\section*{Acknowledgement}

We acknowledge fruitful discussions with Alain Brizard, Phil Morrison, Cristel Chandre, Bastien Fernandez and the \'Equipe de Dynamique Nonlin\'eaire of the Centre de Physique Th\'eorique of Marseille. 

We acknowledge financial support from the Agence Nationale de la Recherche (ANR GYPSI). This work was also supported by the European Community under the contract of Association between EURATOM, CEA, and the French Research Federation for fusion study. The views and opinions expressed herein do not necessarily reflect those of the European Commission.

\end{document}